\documentclass[%
 reprint,
 amsmath,amssymb,
 aps,
 pre,
]{revtex4-2}

\usepackage{dcolumn}
\usepackage{bm}

\usepackage[utf8]{inputenc}
\usepackage[T1]{fontenc}
\usepackage{mathptmx}
\usepackage{etoolbox}
\usepackage[dvipdfmx]{graphicx}
\usepackage{color}
\usepackage{amsmath}
\usepackage{amsfonts}
\usepackage{amssymb}
\usepackage{comment}
\usepackage{siunitx}
\usepackage{here}

\begin{document}

\preprint{APS/123-QED}

\title{Deformation and motion of giant unilamellar vesicles loaded with gold nanoparticles driven by induced charge electro-osmotic flow}

\author{Kotaro Nakazawa}
\affiliation{Department of Applied Physics, Tokyo University of Science, 6-3-1 Nijuku, Katsushika-ku, Tokyo, 125-8585, Japan}
\author{Yoshinori Sonoyama}
\affiliation{Department of Applied Physics, Tokyo University of Science, 6-3-1 Nijuku, Katsushika-ku, Tokyo, 125-8585, Japan}%
\author{Yutaka Sumino}
\email{ysumino@rs.tus.ac.jp}
\affiliation{Department of Applied Physics, Tokyo University of Science, 6-3-1 Nijuku, Katsushika-ku, Tokyo, 125-8585, Japan}%
\affiliation{
 Water Frontier Science \& Technology Research Center, and Division of Colloid Interface, Research Institute for Science \& Technology, Tokyo University of Science, 6-3-1 Nijuku, Katsushika-ku, Tokyo, 125-8585, Japan
}%

\date{\today}

\begin{abstract}
A vesicle is a spherical structure composed of a phospholipid bilayer that is used as a container for chemicals, both {\it in vivo} and {\it in vitro} systems. In both cases, the vesicles can be passively moved using external molecular motors or flows. The active motion of the vesicles can potentially expand their applications in microfluid devices. In this study, we created giant unilamellar vesicles (GUVs) that loads dodecanethiol-functionalized gold nanoparticles (AuNPs) using natural swelling method. An external alternating current (AC) electric field was applied to the sample to drive the system. A flow was confirmed with dense optical flow method around GUVs, even in the absence of AuNPs. The quadratic dependence of the flow on applied elecric fields confirms that the flow is due to induced charge electro-osmotic (ICEO) mechanism. Furthermore, the GUVs containing AuNPs moved and deformed significantly under external AC electric fields compared with those without AuNPs. We also confirmed that the translational speed of GUVs was positively correlated with the volume ratio of AuNPs. These experimental results suggest that the motion and deformation of GUVs were cause by ICEO flow, which was unbalanced owing to the presence of localized AuNPs on the membrane.
\end{abstract}

\maketitle


\section{Introduction}
Cellular membrane composed of a phospholipid bilayer separates interior and exterior of a cell. Such a structure is called a vesicle, which also participates in the storage and transportation of biological chemicals and enzymes within a cell~\cite{Alberts2022, Phillips2012, Gozen2022}. Particularly, when the bilayer is single and the radius is larger than 1-10 \SI{}{\micro \meter}, the vesicle is called a giant unilamellar vesicle (GUV). GUVs are used as a model system for cells and organelles and are particularly relevant in {\it in vitro} experiments, for example, as a body component of a micro-robot~\cite{Li2023, Shoji2007}, a transporter of drug-delivery systems~\cite{Alavi2017,Liu2021b, VanderKoog2022,Daraee2016,Guimaraes2021}, and a container for microreactors~\cite{Walde2010, Elani2016}. Typically, GUVs are driven by external forces or flows. However, such external control of each GUV is neither efficient nor practical for assembling GUVs into organized devices. For the efficiency of these applications, the active motion of GUVs is desirable so that functions can be performed in a bottom-up manner.

Recent progress in molecular design manipulation has enabled the introduction of self-propulsion to GUVs. Examples include GUVs driven by hapotaxis~\cite{Solon2006a}, ion exhange~\cite{Miura2010, Kodama2016}, and redox reactions~\cite{Wilson2012}. Drug-delivery of anti-cancer medicine was proposed~\cite{Abdelmohsen2016, Hortelao2018} as a potential application for motile GUVs. These previous work emphasizes the advantage of self-propelled GUVs as a chemical container.

This study focuses on the techniques used to create self-propelled particles for driving the GUVs with propulsion at an individual level thorugh the application of homogeneous AC external electric fields. These techniques have been demonstrated to induce the motion of self-propelled particles~\cite{Marchetti2013, Gangwal2008, Nishiguchi2015, Iwasawa2021}. AC electric fields provide a continuous energy supply to the particles, and the frequency and strength of the applied fields can be adjusted to control the individual motion of each particle.

\begin{figure}
\centering
\includegraphics{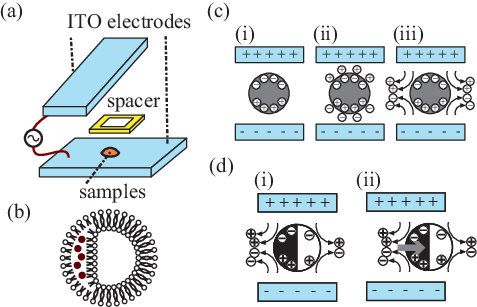}
\caption{\label{fig:figure_1} (a) Schematic of an experimental setup. Samples were placed between ITO-coated electrodes whose gap was controlled by a spacer of 25 or 50 \SI{}{\micro \meter} thickness. An AC electric field was applied to the sample parallel to the direction of observation. (b) Schematic of the AuNP-loaded GUV. AuNPs were loaded in the bilayer. It is suggested that the AuNPs are accumulated owing to the curvature effect of the bilayer. (c) Schematic for the ICEO flow. (i) When an electric field is applied to the sample, an image charge is induced on the GUV's surface owing to the contrast in the dielectric constant. (ii) An electric double layer is forced to screen induced charges. (iii) Electric field tangential to the surface of the GUV that drives the electric double layer to create ICEO flow. (d) Assumed mechanisms to propel the GUV by an asymmetric ICEO flow. When the surface is asymmetric, ICEO flow is also asymmetric to propel GUV, as shown in (i) and (ii).}
\end{figure}

When an external electric field is applied using the setup shown in Fig.~\ref{fig:figure_1}(a), for objects dispersed in the aqueous phase, image charges are exerted on the object surface owing to the contrast of the dielectric constant, as shown in Fig.~\ref{fig:figure_1}(c-i). These image charges created an electric double layer around the object(Fig.~\ref{fig:figure_1} (c) (ii)). Further, the electric fields are distorted to produce components tangential to the surface of the objects that penetrate the electric double layer. Consequently, a flow appears on the surface of the object, which is the so-called induced charge electro-osmotic flow (ICEO flow)(Fig.~\ref{fig:figure_1} (c) (iii)).~\cite{Bazant2004, Squires2004, Gangwal2008, Bazant2009}. If an object is symmetrical, the created ICEO flow is azimuthally symmetrical and does not create a motion perpendicular to the applied electric field. However, if an object has an anisotropic shape or surface, such as a Janus particle~\cite{Zhang2017, Nishiguchi2015, Iwasawa2021}, the created flow becomes asymmetric, inducing motion perpendicular to the applied electric fields(Fig.~\ref{fig:figure_1}(d)). Janus particles under AC electric fields are prototype systems for studying the collective motion of self-propelled particles and active matter. 

In this study, we created self-propelling vesicles under AC electric fields. Gold nanoparticles (AuNPs) were loaded in the bilayer of GUVs using natural swelling methods. According to previous studies~\cite{Rasch2010}, such loaded AuNPs tend to accumulate on a bilayer membrane to reduce the excess elastic energy owing to distortion (Fig.~\ref{fig:figure_1}(b)). It was reported that vesicles with a diameter of 10 \ SI{}{\ nano \ meter} can obtain Janus structure by adding AuNPs. Here, we expect GUVs loaded with AuNPs to create similar asymmetric surfaces like Janus particles spontaneously, and to be propelled under external AC electric fields.

\section{Experiments}
GUV was prepared using 1,2-dioleoyl-sn-glycero-3-phosphocholine, DOPC (Avanti 850375), and fluorescent lipid, rhodamine B, 1,2-dihexadecanoyl-sn-glycero-3-phosphoethanolamine, triethylammonium salt, rhodamine DHPE (Invitrogen L1392). Chloroform was purchased from Fujifilm Wako Pure Chemical Co. (035-02616) and used to prepare GUVs. We used dodecanethiol functionalized gold nanoparticles (AuNPs) purchased from Sigma-Aldrich (660434). The AuNPs were 3-5 \SI{}{\nano \meter} in size, and the solution was 2 \% (w/v) in toluene. To visualize the surrounding fluid, we used polystyrene (PS) particles with a diameter of 1 \SI{}{\micro \meter} purchased from Polyscience (Fluoresbrite carboxylate microspheres (2.5\% Solids-Late x) 1.0\textmu m YO; 18449). Pure water was prepared using a Millipore MilliQ system.

GUVs were prepared using natural swelling methods~\cite{Reeves1969, Tsumoto2009}. Five \SI{}{\micro \litre} of 10 \SI{}{\milli \mole \per \liter} and 0.1 \SI{}{\milli \mole \per \liter} chloroform solutions of DOPC and rhodamine DHPE were cast onto the inner wall of the glass tube. After evaporating the chloroform in a desiccator, we placed 1 \SI{}{\milli \liter} of pure water and swelled the film for an hour to obtain GUVs at 50 \SI{}{\celsius}. To visualize the flow around the vesicles, we placed PS particles in pure water to swell GUVs.

To load AuNPs, we initially placed AuNP toluene solution with the desired amount into a glass tube, after which the toluene was evaporated. We followed the same protocol as described for the GUVs: addition of a chloroform solution of lipids, evaporation of chloroform, and swelling in pure water. The volume fraction of AuNPs $\phi$ in the lipid was varied as a parameter, where
\begin{align}
\phi = \frac{v_\mathrm{AuNP}}{v_\mathrm{DOPC} + v_\mathrm{AuNP}}.
\end{align}
Here, we obtained $v_\mathrm{DOPC}$ by assuming that the molecular volume of DOPC was 1,304 \SI{}{\cubic \angstrom }~\cite{Murugova2014}. $v_\mathrm{AuNP}$ was estimated based on the mass density of gold as 1.932 $\times$10$^4$ \SI{}{\kilo \gram \per \cubic \meter}.

The chamber consisted of two parallel glass plates coated with indium tin oxide (ITO). The bottom plate was ITO-coated cover slips with thicknesses of 0.13-0.17 \SI{}{\milli \meter} purchased from Alliance Bio (06494). The top plate was an ITO-coated glass plate purchased from Geomatech Co., Ltd. (1007). 
To prevent the adhesion of GUVs to the surface, the plates were immersed in 10 wt. \% pluronic solution (AnaSpec, Inc.,cat. number AS-84042) for 10 min, and then rinsed with pure water.

We used a Kapton film (Dupont,100H) with 25 $\SI{}{\micro\meter}$ thickness as a spacer to create the chamber, as shown in Fig.~\ref{fig:figure_1}(a). The gap width $d$ was assumed to be 25 or 50 $\SI{}{\micro\meter}$. An AC electric field was applied using a function generator (NF WF1968) with $f=$ 10 \SI{}{ \kilo \hertz }. We changed the applied voltage $V_{\mathrm{pp}}$ and estimated the electric field using $E_\mathrm{pp}=V_\mathrm{pp}/d$. The observation was conducted with a digital video camera(Olympus, DP72) installed on an optical microscope(Olympus, IX71) with an objective lens (Olympus, UPLFLN40X and UCPLFLN20X).

\begin{figure}
\centering
\includegraphics[width=0.48\textwidth]{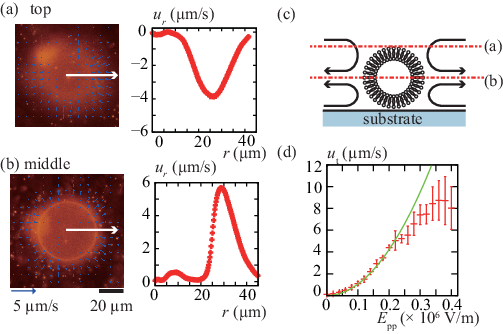}
\caption{\label{fig:figure_2} Observation of ICEO flow induced by the GUV without AuNPs. AC electric fields with $f = $ 10 \SI{}{\kilo \hertz} was applied. The gap between electrodes was 50 \SI{}{\micro \meter}. Radial components of ICEO flow $u_r$ with $E_{\mathrm{pp}}$ = 0.4$\times 10^6$ \SI{}{\volt \per \meter} 
at (a) the top and (b) the middle of the GUV. A flow around the GUV was observed with optical flow methods with the aid of PS particles. $u_r$ was obtained through temporal averaging over 2 s interval with 15 Hz image acquisition and subsequent averaging in the angular direction. Accumulating flow was observed at the top, while diverging flow was observed at the middle. (b) Observed flow intensity $u_r$ at the middle part of the GUV. (c) Position of the images of (a) and (b). (d) Dependence of ICEO flow $u_{\mathrm{t}}$ on electric fields $E_{\mathrm{pp}}$. Here, we obtained the ICEO flow at the midpoint of the four independent vesicles as in (b). The data $u_r$ was averaged for 5 s with 15 Hz image acquisition and subsequently averaged in the angular direction. The averaged peak values $u_{\mathrm{t}}$ over four independent measurements of $u_r$ were plotted against applied electric fields $E_{\mathrm{pp}}$. The error bar represents the standard deviation of four measurements. The green line represents fitting upto $E_{\mathrm{pp}}=0.25\times 10^6$ \SI{}{\volt \per \meter} with theoretical values for ICEO, $u_{\mathrm{t}}=\alpha E_{\mathrm{pp}}^2$, where $\alpha=105 \pm 5$ \SI{}{\micro \meter}$^3$s$^{-1}$ V$^{-2}$. }
\end{figure}

\section{Results and Discussion}

\begin{figure*}
\centering
\includegraphics{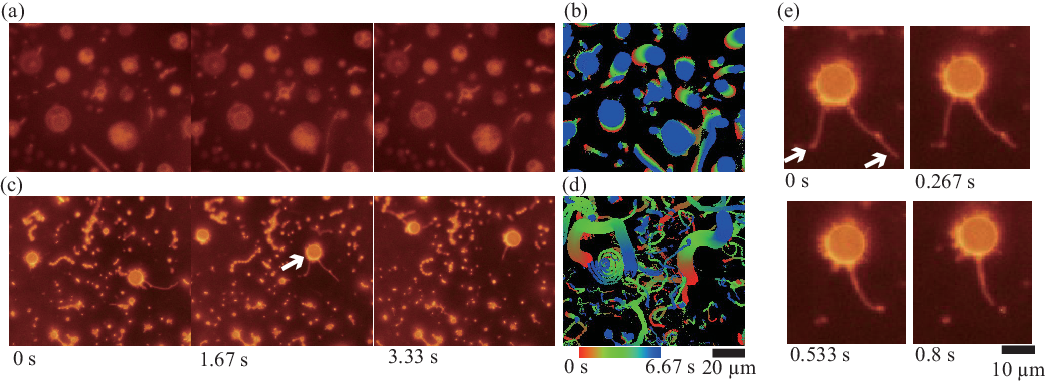}
\caption{\label{fig:fig3} Motion of GUVs under AC electric fields $E_{\mathrm{pp}}$ = 0.8$\times 10^6$ \SI{}{\volt \per \meter} and $f=$ 10 \SI{}{\kilo \hertz} with the gap width of 25 \SI{}{\micro \meter}. (a,b) without (c,d,e) with AuNPs. Despite the assumed existence of ICEO flow, GUVs did not show extensive motion without AuNPs. The difference was distinctive when the motion was compared with the GUVs with AuNPs. GUVs showed filamentous extrudes enlarged in (e) as well as extensive motion, as shown in (b) and (d).  
}
\end{figure*}

To confirm the presence of ICEO flow around the vesicles dispersed in the aqueous phase, we observed the flow around the vesicles using PS particles. It is important to note that GUVs without AuNPs were utilized in this observation. The results are presented in Fig.~\ref{fig:figure_2}(a) and (b). We applied $E_\mathrm{pp}=0.2$×$10^6$ \SI{}{\volt \per \meter} and observed the flow at (a) the top and (b) the middle of the vesicles. The dense optical flow method was used to obtain the flow fields. In both cases, the flow around the vesicle was axisymmetric, and no significant deformation or motion of the vesicles was observed. The radial component of the flow was averaged over time, and the azimuthal angle was used to obtain $u_r(r)$. $u_r(r)<0$ at the top part indicates convergent flow to the vesicles, whereas $u_r(r)>0$ in the middle part indicates divergent flow, as shown in Fig.~\ref{fig:figure_2}(c). We obtained the peak values of $u_r(r)$ at the middle of the vesicle as $u_{\mathrm{t}}(E_\mathrm{pp})$.

The dependence $u_{\mathrm{t}}(E_\mathrm{pp})$ on $E_\mathrm{pp}$ was measured, as shown in Fig.~\ref{fig:figure_2}(d). We confirmed that the flow was proportional to twice the electric field, as discussed for other ICEO flow~\cite{Bazant2004, Squires2004, Gangwal2008, Bazant2009}. We fitted the result with $u_{\mathrm{t}}=\alpha E_\mathrm{pp}^2$ and obtained $\alpha=105 \pm 5$ \SI{}{\cubic \micro \meter \per \second \per \square \volt}

The obtained $\alpha$ was 10$^{-2}$ for the flow around chrome-coated Janus particles made of polystyrene~\cite{Gangwal2008}. Considering the thinness of the lipid membrane and the relatively small contrast of the dielectric constant, the small $\alpha$ compared to that in the previous study is still consistent with the ICEO flow mechanism.

\begin{figure}
\centering
\includegraphics[width=0.48\textwidth]{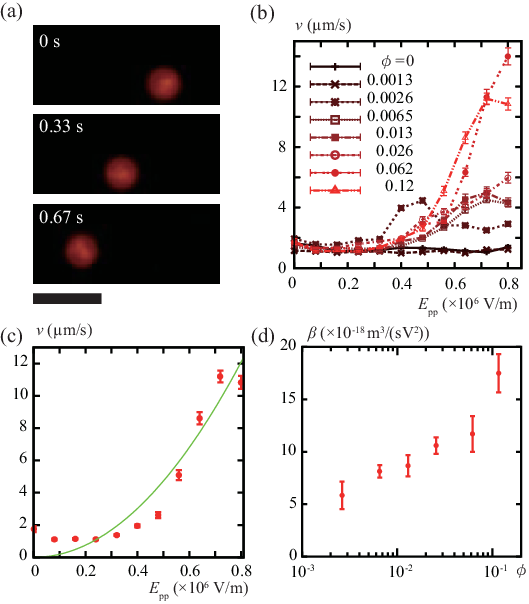}
\caption{\label{fig:fig4} Quantitative measurements of motions of GUVs. The data with GUVs whose diameter was close to 10\SI{}{\micro \meter} were used. Electric fields were $f=$ 10 \SI{}{\kilo \hertz} and the gap width was 25 \SI{}{\micro \meter}. 
(a) Typical motion of a GUV under AC electric fields $E_{\mathrm{pp}}$ = 0.8$\times 10^6$ \SI{}{\volt \per \meter}. Scale bar: 20 \SI{}{\micro \meter}. (b) Observed speed $v$ of the GUVs. We varied the concentration of AuNPs in the seed solution $\phi$ and the strength of electric fields $E_{\mathrm{pp}}$. (c) Fitting of $v$ when $\phi=0.12$ with $v = \beta(\phi) E_{\mathrm{pp}}^2$. (d) Obtained fitting parameter $\beta(\phi)$ shows a logarithmic increase with the increase of AuNP ratio, $\phi$.}
\end{figure}

We then observed the behavior of the GUV loaded with AuNPs in the membrane. Typical snapshots of the GUVs with and without AuNPs are shown in Fig.~\ref{fig:fig3} (a) and (c), and their superimposed images are shown in Fig.~\ref{fig:fig3} (b) and (d). The images in Fig.~\ref{fig:fig3} (c), (d), and (e) corresponds to the data for GUV with $\phi=0.12$. Electric fields of $E_\mathrm{pp}=0.8 \times 10^6$ \SI{}{\volt \per \meter} and $f=$ 10 \SI{}{\kilo \hertz} were applied with a gap width of 25 \SI{}{\micro \meter}. We observed a small random displacement of GUVs, even without AuNPs, reflecting the formation of an ICEO flow around the GUVs. However, the motion was significantly enhanced in the presence of the loaded AuNPs.

In addition to translational motion, we observed the deformation of GUVs characterized by filamentous extrusion, as shown in Fig.~\ref{fig:fig3} (c), as indicated by the white arrows. This filamentous deformation repeatedly extended and suddenly retreated, as shown in Fig.~\ref{fig:fig3} (e). These observations suggest that the inhomogeneous distribution of AuNPs on the membrane induces an asymmetric flow that propels and deforms GUVs.

To quantify the effect of AuNPs, we focused on the translational motion of the vesicles, as shown in Fig.~\ref{fig:fig4}(a). We observed the translational speed $v$ of GUVs with diameters of approximately 10 \SI{}{\micro \meter} while changing the volume ratio of AuNPs $\phi$. We also changed $E_\mathrm{pp}$ as a parameter. The obtained data are shown in Fig.~\ref{fig:fig4}(b). The speed $v$ also fits well with the square law $v=\beta(\phi=0.12) E_\mathrm{pp}^2$, as shown in Fig.~\ref{fig:fig4}(c). The fitting revealed that the obtained coefficient $\beta(\phi)$ was positively correlated with $\phi$, as shown in Fig.~\ref{fig:fig4}(d). This proves that the addition of AuNPs enhances the motion of GUVs. The inclusion of AuNPs in the membrane of the GUV was indirectly confirmed by these data. Furthermore, we observed that the dependency of $\beta$ on $\phi$ is logarithmic, as shown in Figs.~\ref{fig:fig4}(d). This fact may suggest some physical implications for ICEO flow and/or the distribution of AuNPs; however, we presently do not have clear mechanisms.

\section{Conclusion}
In this study, we observed the flow generated around GUVs under AC electric fields without loading AuNPs. Using optical flow methods, we observed that the flow velocity was proportional to the square of the applied electric field. This result suggests that ICEO mechanisms create the flow. Interestingly, we observed enhanced membrane fluctuations and GUV motion even in the absence of AuNPs. Such membrane fluctuations were observed for multilamellar vesicles, and the motion was intense in distorted vesicles with lipid aggregates. 
These findings also suggest that the asymmetry imposed on GUV distorts the symmetric ICEO flow, creating motion in the membranes and GUVs. Notably, these fluctuations and motions were much more prominent in the case of GUVs with AuNPs.

The addition of AuNPs led to broken symmetry, reflecting a lateral inhomogeneous distribution on the membrane of the GUV. The GUVs with AuNPs were efficiently propelled when external AC electric fields were applied. Some GUVs also exhibited significant deformation characterized by filamentous extrusions. The speed of motion of the GUVs increased with an increase in the electric field, and the volume ratio of loaded AuNPs increased. The speed was approximately slower than that of the ICEO flow observed around the GUV without AuNPs. The role of AuNPs was to break the spherical symmetry of GUVs and create unbalanced flow fields to propel them. 

Our observations show that the typical speed of ICEO flow was approximately 10 \SI{}{\micro \meter \per \second}. These values are smaller than those reported for Janus particles ($\sim$ a few 10 \SI{}{\micro \meter \per \second}~\cite{Gangwal2008}) made of metal-coated dielectric materials. The observed value of ICEO flow around the thin membrane of GUVs was smaller than that of Janus particles and remained consistent. However, further theoretical validation is required. 

Unfortunately, the distribution of AuNPs on the membrane has not been confirmed yet. A previous study~\cite{Rasch2010} confirmed this using cryo-TEM with small unilamellar vesicles with a typical diameter of 50 \SI{}{\nano \meter}. Cryo-TEM is not applicable to visualize the dynamic motion observed in our system.
In this study, we observed the dependence of $\beta$, which symbolizes the motion of GUVs on $\phi$ in a logarithmic manner. The positive correlation between $\beta$ and $\phi$ clearly shows that the inclusion of AuNPs leads to the motion of GUVs through their asymmetric distribution on the surface. However, the cause of the logarithmic dependence remains unclear. This may reflect the aggregate size and shape of AuNPs on the membrane and/or the induced ICEO flow. This analysis requires further theoretical consideration and should be considered in future studies.
We also found that the yield of GUVs with AuNPs was low when $\phi$ was greater than 0.1, indicating that the stability of the membrane was lost because of the excessive amount of AuNPs within the membrane. 
These difficulties can be resolved by directly observing AuNPs on the membrane. 
Such visualization of AuNPs could also help predict and control the motion and deformation of GUVs with AuNPs.

For the biomedical and chemical applications, the drawback of the present technique is the need for high-strength electric fields to induce motility in GUVs. Consequently, our technique may be relevant under artificial conditions rather than {\it in vivo} conditions. One important application of the results of this study is the use of ICEO-driven GUVs as a flexible and motile conteiners for micro-chemical laboratories. GUVs can transport contents while simultaneously separating them from external medium. It is noteworthy that GUVs can be fused by irradiating laser light to the AuNPs embedeed in a membrane~\cite{Rorvig-Lund2015}. Thus, combining such a technique with the presented results extends the potential use of GUVs as motile microreactors. For such purposes, additional external control of the direction of motion may be desirable, and aid of laser light may be useful.

From a theoretical and academic perspective, the relationship between membrane deformation and the position of AuNPs is of great interest. A recent study suggested the accumulation of self-propelled particles with a deformed membrane~\cite{Nikola2016, Vutukuri2020, Kokot2022}. Therefore, our system may exhibit similar interactions; however, the visualization of AuNPs is an essential technique. Such studies will be of interest to researchers in active matter physics.


\begin{acknowledgments}
This work was partially supported by JSPS KAKENHI Grant Numbers JP16K13866, JP16H06478, JP19H05403, 21H00409 and 21H01004. This work was also supported by JSPS and PAN under the Japan-Poland Research Cooperative Program ``Spatio-temporal patterns of elements driven by self-generated, geometrically constrained flows,'' and the Cooperative Research of ``Network Joint Research Center for Materials and Devices'' with Hokkaido University (Nos.~20161033, 20171033, 20181048). We would like to thank Editage (www.editage.jp) for English language editing.
\end{acknowledgments}

K. Nakazawa and Y. Sonoyama contributed equally to this work.

\end{document}